\RequirePackage{fix-cm}
\documentclass[twocolumn]{svjour3}          % twocolumn
\smartqed  % flush right qed marks, e.g. at end of proof
\pdfoutput=1
\usepackage{graphicx}
\usepackage{mathptmx}      % use Times fonts if available on your TeX system
\usepackage{subfigure}
\usepackage{amsmath}
\usepackage{tikz}
\usepackage{pgfplots}
\usepackage{natbib}
\usepackage{multirow}
\usepackage{stfloats}
\usepackage{float}
\usepackage[utf8]{inputenc}   % 允许UTF-8编码
\usepackage[T1]{fontenc}      % 使用T1字体编码
\usepackage{CJKutf8}

\journalname{Microfluidics and Nanofluidics}
\usepackage[font=bf,textfont=md]{caption}

\journalname{Microfluidics and Nanofluidics}
\begin{document}
\begin{CJK}{UTF8}{gbsn}
\begin{sloppypar}
\title{Study on Dynamic Solidification of Digital Droplets and Random Behaviors during the Recalescence Process in a Spiral-shaped Milli-reactor%\thanks{Grants or other notes
}
\titlerunning{Study on Dynamic Solidification of Digital Droplets}        % if too long for running head
\author{Yulin Wang \textsuperscript{1} \and Z. L. Wang\textsuperscript{1}*
}
\institute{*Z. L. Wang\\
wng\_zh@i.shu.edu.cn
 \\
\at
\textsuperscript{1} Shanghai Institute of Applied Mathematics and Mechanics, Shanghai Key Laboratory of Mechanics in Energy Engineering, Shanghai University, No.149 Yanchang Road, Shanghai 200072, P. R. China\\
Tel.: +86-21-56331451\\
Fax: +86-21-56331453\\
%  \\
%\and
%          \
}
\date{Received: 28 November 2024 / Accepted: date}
\maketitle % need full-width title
\begin{abstract}
The freezing of droplets is a complex interdisciplinary research topic involving physics, chemistry, and computational science. This phenomenon has attracted considerable attention due to its significant applications in aerospace, meteorology, materials science, cryobiology, and pharmaceutical development. The development of microfluidic technology provides an ideal platform for microscopic physical research. 
In this study, we designed a spiral-shaped milli-reactor with a T-junction microchannel to generate digital droplets for studying and observing the digital freezing process of droplets. During the study of the recalescence and solidification processes of digital droplets dynamically moving in microchannels, we found that although the digital generation of droplets in our channel aligns well with the literature, achieving the digitalization of the droplet freezing process is very challenging. Even the initial phase of freezing (the recalescence process) exhibits significant randomness.  
A key feature of the randomness in the freezing process is the nucleation position of droplets within the channel, which significantly impacts the digital characteristics and hinders digital freezing. During the investigation of freezing randomness, we identified five distinct nucleation profiles, which largely determine the evolution of the freezing front and the duration of the recalescence phase. However, upon studying the motion velocity of the freezing front, we found that these velocities are temperature-dependent. This aligns with the results of our phase-field simulations and experimental findings, indicating that the release of latent heat during the recalescence process is stable.
Additionally, the randomness in freezing may also stem from the deformation of droplets during the solidification process. In this study, we identified two distinct solidification modes during the freezing phase: one initiating from the droplet's head or tail and the other starting from the middle, with the latter causing significant droplet deformation. Through statistical analysis, we further explored the influence of flow rate variation on the digital clustering of droplet freezing and discovered flow rate parameters that optimize freezing digitalization. For instance, when the oil phase flow rate is fixed, varying the water phase flow rate initially increases and then decreases the flatness factor, reaching a maximum at a water phase flow rate of  \( Q_w = 0.5 \, \text{mL/min} \), indicating optimal clustering of droplets. The findings of this study provide new perspectives and approaches for controlling droplet freezing in microfluidic systems, while also offering significant insights into the unique behaviors and phenomena of nucleation and solidification processes at the microscale.
\end{abstract}

\section{Introduction}
\label{intro}

The freezing behavior of droplets has garnered widespread attention across multiple disciplines, including atmospheric 
\begin{table}[ht]
    \centering
    \begin{tabular}{|p{15pt}p{200pt}|} % 调整列宽，根据需要修改
        \hline % 第一条横线
        \multicolumn{2}{|l|}{\textbf{Nomenclature}} \\[5pt]
        
        T & Temperature,~℃ \\[0pt]
        t & Time,~s \\[0pt]
        $L_{d}$ & Droplet length,~mm \\[0pt]
        $Q_{w}$ & Water flow rate,~mL/min \\[0pt]
        $Q_{O}$ & Silicone oil flow rate,~mL/min \\[0pt]
        $v_{g}$ & Growth velocity of ice crystal,~m/s \\[0pt]
        P & Relative position of the ice-water interface,~mm \\[0pt]
        $D_{p}$ & The position of the droplet as it nucleates in the spiral tube,~mm \\[0pt]
        $\Delta G$ & The work required to form an ice nucleus \\[0pt]
        $\Delta \mu$ & The difference of the chemical potential between the liquid phase and the crystal \\[0pt]
        $\Sigma$ & The water-ice interfacial tension \\[0pt]
        A & The surface area per water molecule \\[0pt]
        $T_{m}$ & Melting temperature,~K \\[0pt]
        $N_{n}$ & Number of droplets nucleated at a given location in the pipe \\[0pt]
        L & Latent heat of fusion,~$J/cm^{3}$ \\[0pt]
        $c_{p}$ & Specific heat,~$J/cm^{3}K$ \\[0pt]
        $\kappa$ & Thermal diffusion coefficient,~$cm^{2}/s$ \\[0pt]
        $\mu$ & Interface mobility,~$cm/(Ks)$ \\[0pt]
        $\sigma$ & Interface energy,~$J/cm^{2}$ \\[0pt]
        $\omega$ & Interface thickness,~cm \\[0pt]
        $\alpha$ & Interface freedom \\[0pt]
        m & Phase field mobility \\[0pt]
        $\overline{\epsilon}$ & Computational parameter \\[0pt]
        $\epsilon_{k}$ & Anisotropy constant \\[0pt]
        $\Omega$ & Dimensionless undercooling \\[0pt]
        dx & Dimensionless grid spacing \\[0pt]
        dy & Dimensionless grid spacing \\[0pt]
        dt & Dimensionless time \\[3pt]
        \hline % 第四条横线
    \end{tabular}
    \label{tab:one_column_table}
\end{table}
science (\cite{kalita2023microstructure,knopf2023atmospheric,miller2021development,tarn2020chip,tarn2021homogeneous,addula2024modeling}), cryobiology (\cite{sgro2007thermoelectric,prickett2015effect,powell2020isochoric,consiglio2024review,kangas2024mathematical}), pharmaceuticals (\cite{kasper2011freezing,ge2024research}), and food science (\cite{you2021control,chhabra2024spray}). 
Among these studies, droplet nucleation and freezing are key areas of focus.
Microfluidic devices, with their ability to generate large numbers of monodisperse droplets, high surface-to-volume ratios, and independent control over individual droplets (\cite{teh2008droplet,zhu2017passive}), provide an ideal platform for droplet freezing research due to their rapid development. In the process of droplet freezing, nucleation requires overcoming the free energy barrier (\cite{assegehegn2019importance,prestipino2018barrier,laaksonen2021nucleation}), and metastable supercooled water can persist for extended periods. For instance, cloud droplets in the atmosphere(\cite{demott2011resurgence,tarn2021homogeneous,tarn2020chip,whitesides2006origins}) or ultrapure, particle-free pharmaceutical formulations (\cite{searles2001ice,deck2022stochastic}) exhibit nucleation temperatures that depend on parameters such as the bulk solution volume, cooling rate, and nucleation mode. The free energy barrier, which represents the work required to form an ice nucleus of size \(i\), can be expressed by the following equation (\cite{tanaka2019theoretical}):  
$\Delta G = - \Delta \mu i + \Sigma A i^{2/3}$  
(where $\Delta \mu$ is the difference of the chemical potential between the liquid phase and the crystal, $\Sigma$ is the water-ice interfacial tension, and $A$ is the surface area per water molecule).  
The presence of this free energy barrier leads to the stochastic nature of the droplet nucleation process. Therefore, the solidification process of droplets in microchannels requires further in-depth investigation.

\begin{figure}[!htbp]
\centering
\includegraphics[width=\linewidth]{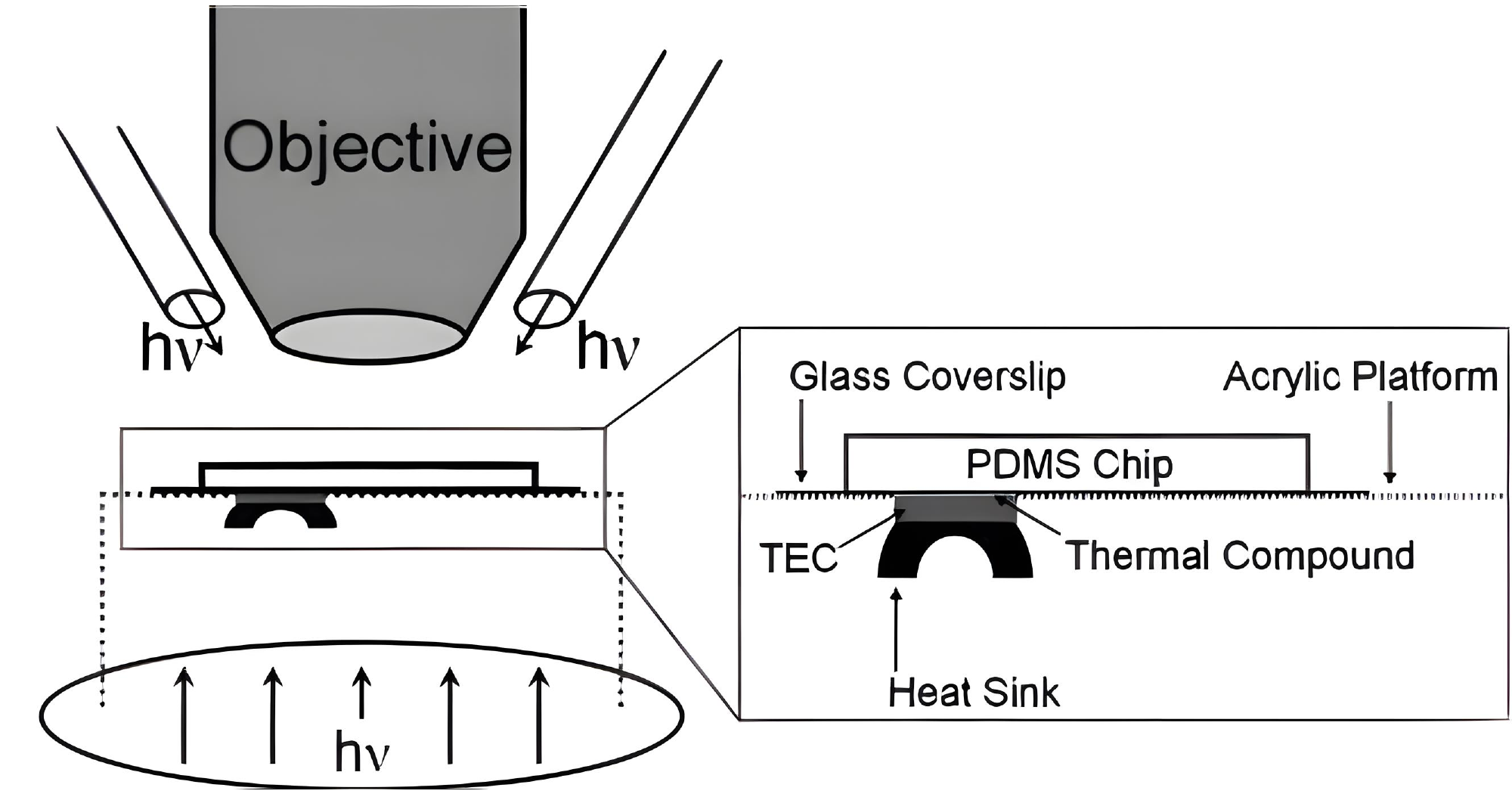}  % Adjusts the image to the width of one column
\caption{The schematic diagram of the Peltier elements is shown, where the TEC section specifically represents the Peltier components(\cite{stan2009microfluidic}).}
\label{fig:1}
\end{figure}

In recent years, microchannel technology has been increasingly applied to the study of droplet freezing, with most research focusing on ice nucleation rates. Stan et al.(\cite{stan2009microfluidic}) developed a platform for ice nucleation experiments by placing a series of Peltier elements on a microfluidic device (as shown in Figure \ref{fig:1}), capable of generating a thermal gradient along the entire length of the chip. As droplets flowed through the channel, they determined the temperature within the channel using a combination of microfabricated thermometers and modeling. The freezing position of the droplets was found to correlate with the temperature at that location.  
By setting the continuous phase flow rate to 3 mL/hr and the dispersed phase flow rate to 0.065 mL/min, they investigated both homogeneous freezing and heterogeneous freezing of water droplets containing silver iodide. This platform demonstrated the potential of microfluidic freezing methods in droplet freezing studies. Using this setup, they observed dendritic growth during the recalescence stage but did not elaborate on the solidification process. Although they statistically analyzed the freezing positions of droplets at different temperatures, their experiments were conducted in channels with varying temperature gradients, leaving the case of a uniform temperature condition unexplored.  
%\label{sec:2}
%Text with citations \citealt{RefB} and \citealt{RefJ}.

Tarn et al.(\cite{tarn2021homogeneous,tarn2020chip}) designed a microfluidic chip with two independent channel structures: a main channel for the generation and freezing of water-in-oil droplets and a reference channel that carried only oil, serving as an alternative to the main channel for measuring the temperature of the flowing oil. Using a water-phase flow rate of 0.05 \textmu L/min and an oil-phase flow rate of 24 \textmu L/min, they investigated homogeneous nucleation of water and calculated the homogeneous nucleation rate coefficient, validating the functionality of their device.  
Shardt et al. (\cite{shardt2022homogeneous}) utilized the Microfluidic Ice Nuclei Counter Zurich to study the homogeneous freezing of droplets of different volumes under varying cooling rates. This research covered a temperature range of 236.5–239.3 K with a precision of 0.2 K. It not only provided precise data but also employed Monte Carlo simulations to analyze the influence of droplet numbers on the extraction of nucleation rates, offering critical insights into the droplet nucleation mechanism.  

Isenrich et al.(\cite{isenrich2022microfluidic}) utilized a PDMS microfluidic platform to generate droplets, which were subsequently directed into a fluoropolymer tube downstream. The generated droplets were stored in a PTFE tube (inner diameter: 0.25 mm, outer diameter: 0.56 mm), and freezing was carried out using an ethanol bath. This setup enabled continuous flow freezing of droplets while allowing repeated freeze-thaw experiments on the same droplet.  
Using this device, they investigated nucleation in both pure water and suspensions containing microcline to simulate the formation of ice nuclei in atmospheric conditions. 

Microfluidic devices have also demonstrated potential applications in the field of biology (\cite{sgro2007thermoelectric,shardt2022homogeneous}). Sgro et al. (\cite{sgro2007thermoelectric}) were among the first to use Peltier elements, placing them above or below the chip to freeze droplets. This device was used for freezing droplets to encapsulate cells, enabling biological applications. The method required only simple cooling of the chip to ensure droplet freezing (with the Peltier element(Figure \ref{fig:1}) set to -20°C) and did not involve measuring the freezing temperature of the droplets. They used COMSOL to simulate the temperature variations of droplets inside the channel. 

In summary, microchannel-based droplet freezing is primarily utilized for ice nucleation studies (\cite{tarn2020chip,tarn2021homogeneous,shardt2022homogeneous,stan2009microfluidic,isenrich2022microfluidic}),focusing mainly on whether nucleation occurs and the corresponding nucleation temperatures, with the aim of investigating ice nucleation rates. These studies are largely applied to the freezing of water droplets in clouds, while the dynamic freezing process of droplets within channels and the distribution patterns of droplet nucleation are often overlooked. The distribution patterns of droplet nucleation significantly influence the digital freezing process of dynamic droplets. Therefore, it is essential to conduct in-depth research into the dynamic solidification behavior of droplets in microchannels.

The Peltier device is commonly used in microchannel droplet freezing experiments. However, it has a relatively small cooling area, and large-area Peltier devices can lead to uneven cooling, making temperature control difficult. Additionally, the Peltier device requires a heat dissipation system, which results in a larger overall device size. Therefore, it is not suitable for use with a spiral-shaped microreactor.

This study aims to investigate the issue of digitalized droplet solidification by designing a spiral-shaped milli-reactor for droplet generation and freezing. The compact structure of the spiral-shaped channel not only facilitates experimental observation but also enables a more detailed examination of different freezing stages. This work focuses on the recalescence and solidification processes of droplets, particularly exploring the nucleation distribution patterns and their controlling factors during the recalescence phase. Notably, no prior studies have applied spiral-shaped channel structures to microchannel droplet freezing experiments, offering new perspectives and methodologies for the related research field.  

\section{Experimental setup}
\begin{figure*}[!htbp]\centering
\includegraphics[width=\linewidth]{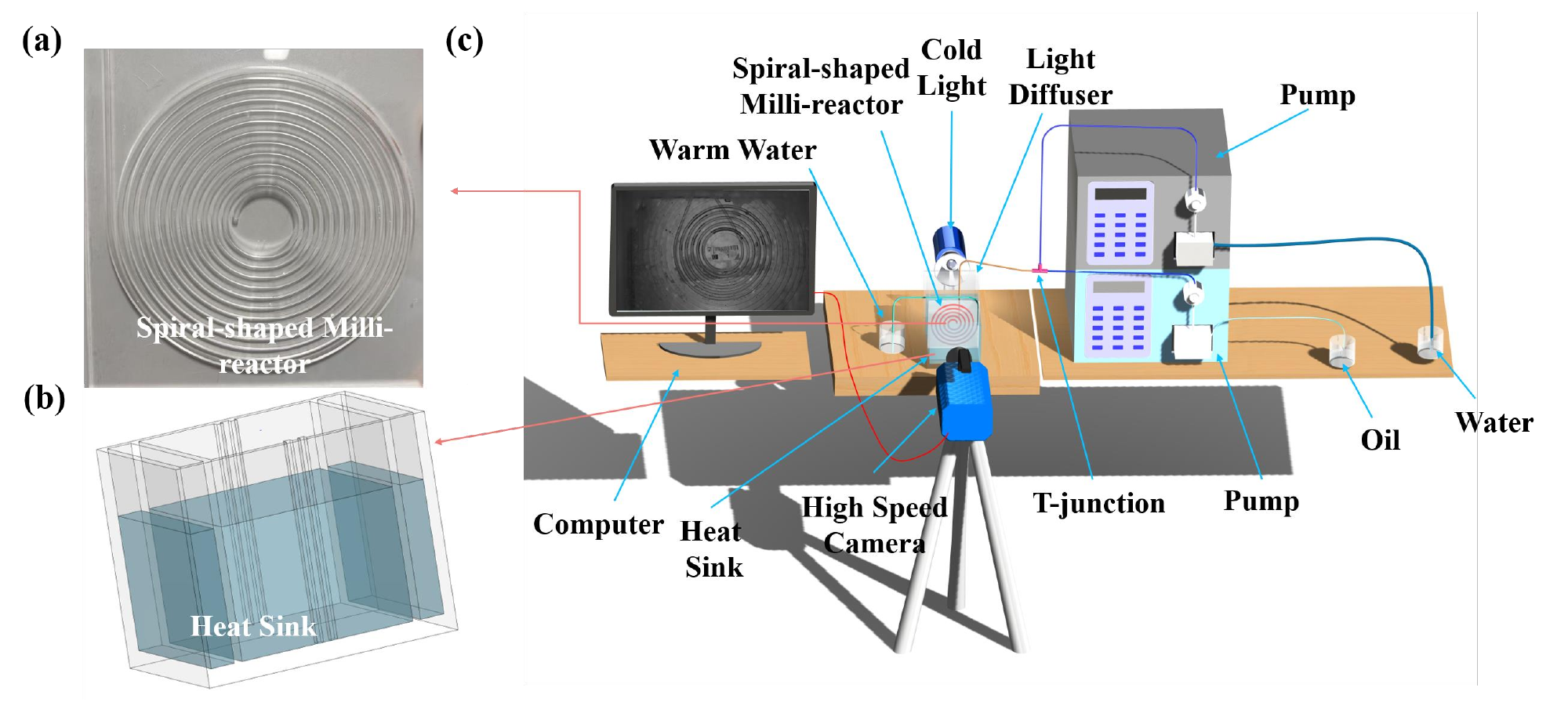}
\caption{Spiral-shaped Milli-reactor for Droplet Freezing: (a) Main components of the Spiral-shaped Milli-reactor; (b) Heat bath; (c) Detailed view of the experimental setup.}
\label{fig:2}
\end{figure*}

In this study, a T-junction was used to generate oil-water two-phase flow droplets, with water serving as the dispersed phase and oil as the continuous phase. Following the T-junction, a spiral-shaped channel and heat sink were implemented for droplet freezing, as shown in Figure \ref{fig:2}.

\subsection{Experimental Setup Design}
\begin{figure}[!htbp]
\centering
\includegraphics[width=\linewidth]{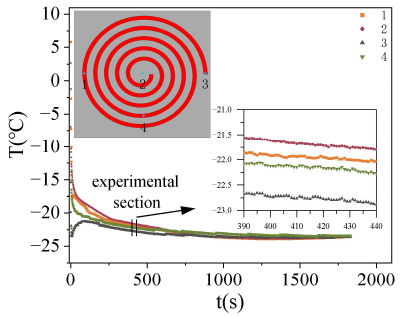}  % Adjusts the image to the width of one column
\caption{Temperature variation at measurement points over time.}
\label{fig:3}
\end{figure}

Figure \ref{fig:2} illustrates the details of the experimental setup. A high-speed camera (Phantom V611-16G-M) equipped with a lens (AT-X M100 PRO D) was used to capture images. For droplet generation, the following equipment and components were utilized: i) a high-pressure infusion pump (SANPTAC, AP0010) and a high-pressure constant flow pump (SANPTAC, SO6010); ii) a T-junction (with an inner diameter of 1 mm for the continuous phase inlet, dispersed phase inlet, and outlet); and iii) a transparent polytetrafluoroethylene (PTFE) tube connected at the outlet(with an inner diameter of 1 mm and an outer diameter of 2 mm). The tube was fixed onto an acrylic plate with Archimedean spiral grooves, as shown in Figure \ref{fig:2}(a). The equation of the spiral groove is $R=11+2/\pi\times\theta$ (mm). 
For droplet cooling, a heat sink was employed: i) the acrylic plate holding the PTFE tube was immersed into it; ii) four thermocouples were fixed at the four positions shown in Figure \ref{fig:3}. To improve image clarity and prevent fogging or frost formation on the surface of the heat sink, alcohol was sprayed on the acrylic box. At the outlet of the channel, frozen droplets often caused blockages. To resolve this, a cup of warm water was placed at the outlet to melt the frozen droplets.

\subsection{Experimental Preparation}

In the experiment, a heat sink was used to cool the spiral-shaped milli-reactor, as shown in Figure \ref{fig:2}(b). Two identical acrylic containers were prepared in advance, each containing 500 mL of anhydrous ethanol (purity: 99.7\%, supplier: Jiangsu Huating Biotechnology Co., Ltd.), 500 mL of glycerol (Sinopharm Group Co., Ltd.), and 500 mL of deionized water. The components were thoroughly mixed and placed in a refrigerator (Xingxing brand, adjustable cooling/freezing chamber) set to freezing mode 3. The mixture was frozen for 24 hours to ensure uniform cooling within the heat sink, thereby minimizing internal temperature gradients.

To achieve droplet freezing, the solidification temperature of the continuous phase material must be significantly lower than that of the dispersed phase. Dimethylsilicone oil is colorless, odorless, and insoluble in water, with a working temperature range of -50°C to 200°C, making it ideal for visualization and low freezing point requirements. Dimethylsilicone oil with a viscosity of 0.00455 Pa$\cdot$s (Dow Corning) was used as the continuous phase material.
The nucleation temperature of pure water can be as low as -35°C, which is difficult to achieve using the heat sink. To lower the energy barrier for droplet nucleation and increase the nucleation temperature, spherical hydrophilic and oleophobic silica powder was used as a nucleating agent, with particle diameters of 20 nm. Before the experiment, the silica powder was prepared as a 0.5 $wt\%$ aqueous solution, allowed to settle for 12 hours, and the supernatant was collected for use.
\subsection{Experimental Procedure}

First, two high-pressure infusion pumps (Sanotac, SP-6015 and AP-0010) were used to transport the supernatant and dimethylsilicone oil (0.00455 Pa$\cdot$s), respectively. The infusion pumps were started to generate stable digitized droplets.  
During this process, the acrylic box, which had been pre-frozen for 24 hours, was removed from the freezer and placed in the designated position. The spiral-shaped milli-reactor was then placed at the center of the acrylic box, and alcohol was sprayed on the observation surface to prevent visual blurring.  
The timer was started, and after 400 seconds, the system reached stability. At this point, the temperature of the pipe and the acrylic plate fixing the pipe stabilized at the heat sink temperature. A high-speed camera (PHANTOM, with an AT-X M100 PRO D lens from Tokina) was used to record the movement and freezing process of the plug-shaped droplets in the middle section of the pipe. The high-speed camera was set to a frame rate of 400 FPS.  

The factors influencing droplet nucleation are complex, with disturbances being one of them. Therefore, directly measuring the temperature of droplets is a common approach to approximate the temperature of the droplets \cite{tarn2020chip, stan2009microfluidic, wu2015heat}. We measured the temperature at four points shown in Figure \ref{fig:3}, and the temperature variation with time is shown in the figure. For a single point, the wall temperature variation does not exceed 0.5°C, and the overall wall temperature change does not exceed 1°C.  
We measured the temperature variation over a period of 30 minutes and found that the temperatures at the four points eventually leveled off and became constant. The temperature gradients between the points were not very large. Since the duration of each experiment was 30 seconds (mainly limited by the high-speed camera's FPS), the temperature variation in our experiments did not exceed 0.5°C. Therefore, we considered this boundary condition to be constant.
\section{Dynamic solidification of droplets}
\label{sec:3}
\subsection{Digital droplet generation}

In the T-type microchannel, the droplet generation modes mainly include slug flow, droplet flow, and parallel flow (\cite{zhao2006liquid}), and these modes depend on the interaction between viscous force, inertial force, and interfacial tension (\cite{zhu2017passive}). Garstecki (\cite{garstecki2006formation}) et al. found that in the case of a low capillary number, the droplet generation is mainly affected by the pressure drop. In the squeezing state, the length of the droplet is determined by the volume ratio of the two immiscible fluids and the following simple scaling law can be obtained:
\begin{equation}
\frac{L}{w} = 1 + \alpha \frac{Q_{\text{in}}}{Q_{\text{out}}}
\end{equation}

\begin{figure}[!htbp]
\centering
\includegraphics[width=\linewidth]{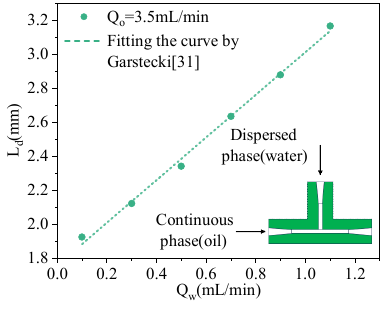}  % Adjusts the image to the width of one column
\caption{Variation of droplet length with the dispersed phase flow rate. The data points represent our experimental results, while the dashed line corresponds to the fitted results based on the scaling law of Refs. (\cite{garstecki2006formation}).}
\label{fig:4}
\end{figure}

As shown in Figure \ref{fig:4}, it is the relationship between the droplet length and the water phase flow rate measured by us. The results show that the droplet length is proportional to the dispersed phase flow rate ratio to the continuous phase flow rate. This further verifies that the experimental results of our T-tube droplet generation are consistent with other studies (\cite{garstecki2006formation}), proving that we can successfully generate digital droplets.

\subsection{ice-water phase front modes in the recalescence stage of droplets}
\label{sec:3}

\begin{figure*}[!htbp]\centering
\includegraphics[width=\linewidth]{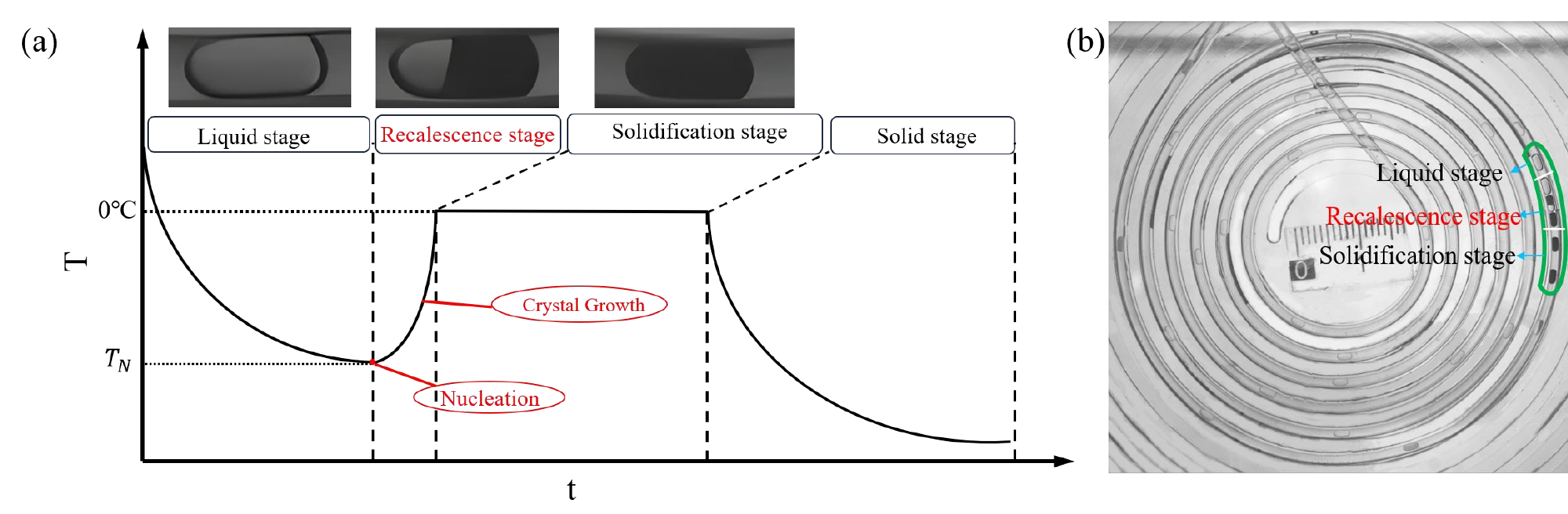}
\caption{Four stages of droplet freezing. (a) The four stages in the entire freezing process of water droplet according to temperature changes; (b)Digital droplet freezing in the spiral micro-reactor. The green circle indicates the state of the same droplet at different stages. The black ones are droplets that have undergone the recalescence stage and entered the solidification phase.}
\label{fig:5}
\end{figure*}

The droplet solidification can be divided into four stages based on temperature changes, as shown in Figure \ref{fig:5}(a). Figure \ref{fig:5}(b) shows the liquid, recalescence, and solidification stages observed in the experiment. (i) Liquid stage: The droplet is cooled from its initial temperature to below \(0^{\circ}C\), while remaining in the liquid state, a condition referred to as supercooled. Despite being below  \(0^{\circ}C\), no ice formes in the supercooled state. (ii)Recalescence stage: When ice nucleates, the droplet enters this stage. The nucleation temperature \(T_{N}\) (not a fixed value) depends on factors such as purity, impurities, and pressure. The degree of supercooling is defined as \(\Delta T = 0 - T_{N}\)℃. The recalescence stage begins with the formation of an ice nucleus inside the droplet, followed by rapid dendritic ice growth, creating an ice-water mixture. In the ice-water mixture region, latent heat release causes the temperature to rise to and stabilize at \(0^{\circ}C\). This stage typically lasts milliseconds, approximately \(10 - 100ms\). (iii)Solidification stage: Following the recalescence stage, ice continues to grow in the form of dense ice. This solidification process is slow, with a duration of about \(1 - 100s\). (iv) Solid stage: The droplet has fully solidified, and its temperature begins to decrease gradually from $0^{\circ}\mathrm{C}$ until reaching thermal equilibrium.
\begin{figure*}[!htbp]\centering
\includegraphics[width=\linewidth]{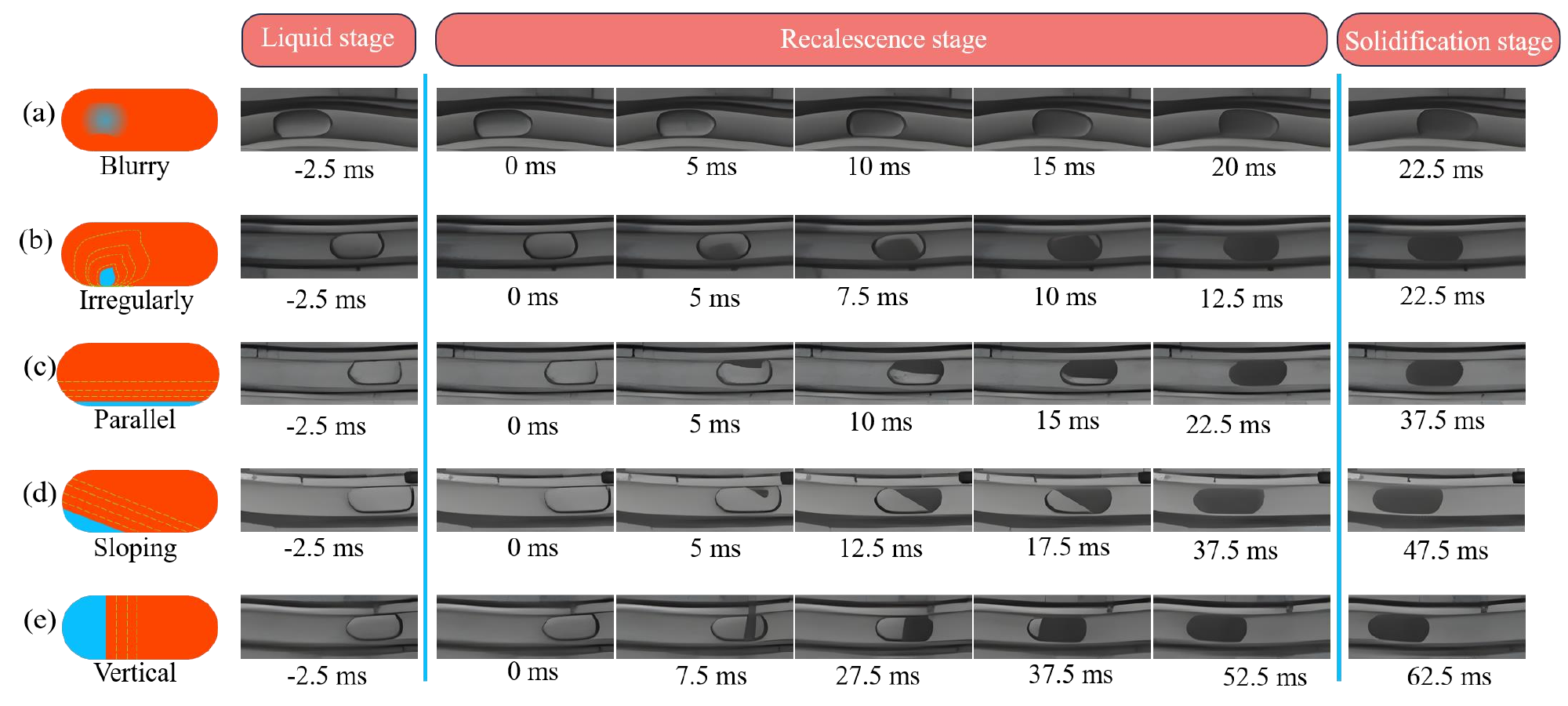}
\caption{Propagation forms of the solid-liquid interface in the recalescence stage. (a) Blurred solid-liquid interface; (b) Clear but irregular interface; (c) Clear interface parallel to the pipeline axis; (d) Clear interface at an angle to the pipeline axis; (e) Clear interface perpendicular to the pipeline axis.}
\label{fig:6}
\end{figure*}

The recalescence stage is a crucial part of the droplet freezing process, during which supercooled water begins to crystallize. However, the short duration of the recalescence phase presents challenges for both experimental observation and theoretical study. Using a high-speed camera, we observed the propagation of the solid-liquid interface in droplets during the recalescence phase. As shown in Figures \ref{fig:6}(a)-(e), different solid-liquid interface propagation patterns were observed for droplets during the recalescence stage. In Figure \ref{fig:6}(a), the interface between the initially formed ice crystal and the liquid water (referred to as the ice front) somewhat blurred. Over time, the ice crystal region darkens and eventually stabilizes, marking the end of the recalescence phase. In Figures \ref{fig:6}(b)-(e), the ice front is more clearly defined: Figure \ref{fig:6}(b) shows an irregular interface with a complex ice front and an irregular ice front shape. The ice front in Figure \ref{fig:6}(c) is parallel to the pipeline axis, and the ice nucleus is formed at the pipeline wall position and propagates in the vertical axis direction. The ice front in Figure \ref{fig:6}(d) forms a certain angle with the pipeline axis, and the propagation direction of the ice front is the vertical direction of the ice front. The ice front in Figure \ref{fig:6}(e) is perpendicular to the pipeline axis, and the ice front propagates along the axis direction. Although there are five forms of ice fronts, the shape of the ice front remains consistent during propagation without any deformation.

\begin{figure}[b!]
\centering
\includegraphics[width=\linewidth]{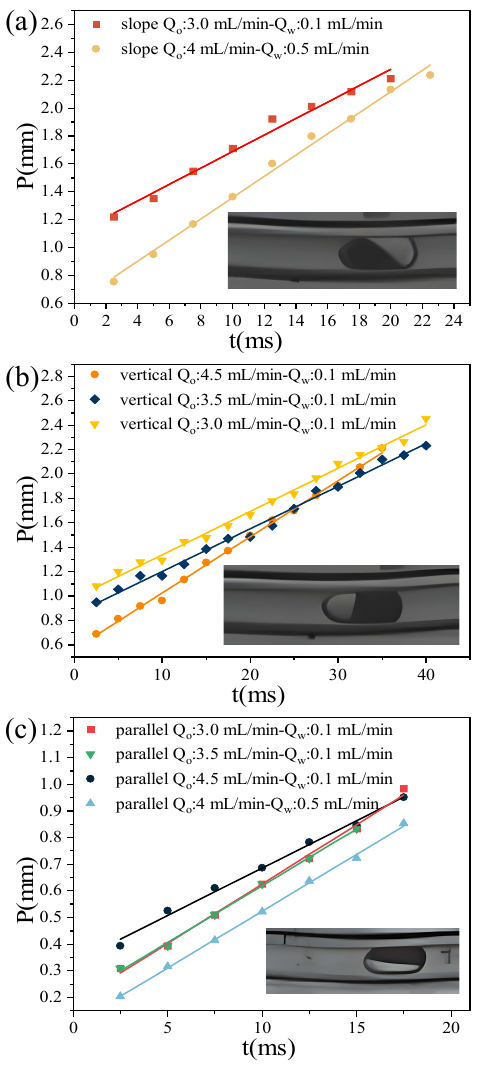}  % Adjusts the image to the width of one column
\caption{Diagrams of the variation of the ice front position with time in the recalescence stage.}
\label{fig:7}
\end{figure}

\begin{figure}[!htbp]
\centering
\includegraphics[width=\linewidth]{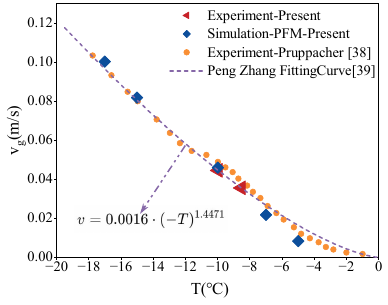}  % Adjusts the image to the width of one column
\caption{Variation of the propagating velocity $v_{g}$ of the ice with the temperature T (℃). The red and blue points represent our experimental results and the phase-field simulation results, respectively. The orange points correspond to the experimental results of Refs.  (\cite{pruppacher1967interpretation}), and the curve represents the fitted results of Refs. (\cite{meng2020dynamic}).}
\label{fig:8}
\end{figure}

In the recalescence stage, ice crystals grow in a dendritic form, and the interface between the dendritic region and the supercooled water (i.e., the ice front) propagates with the growth of the dendrites (\cite{wang2019dendritic}). Figure \ref{fig:7} shows the variation of the ice front position of three ice front profiles in the recalescence stage with time.  
The ice front position is relative. The parallel ice front takes the pipeline wall surface as the reference point and the inclined 
and vertical ice fronts take the tail of the droplet as the reference point. The slope in Figure \ref{fig:7} represents the ice front velocity, which remains constant in our experiments. The phase field method can effectively simulate the growth process of dendrites under the action of the macroscopic field and truly present the profile and evolution of dendrites. By simulating the growth velocity of dendrites, the growth velocity of the ice front can be obtained. The specific phase field model (\cite{wheeler1993computation}), related physical property parameters (\cite{galfre2023phase}), and calculation results are detailed in Appendix A. With the increase of the supercooling degree, the growth velocity of dendrites accelerates, which is mainly related to the release of latent heat at the ice front (\cite{glicksman2004dendritic}).At a low supercooling degree, the temperature gradient of the ice front is gentle, and the heat diffusion is slow, resulting in slow dendrite growth. We calculated the growth velocity of dendrites at different supercooling degrees according to the phase field method and compared it with the experimental results of our experiment and Pruppacher (\cite{pruppacher1967interpretation}), as shown in Figure \ref{fig:8}. The formula in Figure \ref{fig:8} is the fitting relationship between the ice front and the supercooling degree (\cite{meng2020dynamic}), and the detailed formula is as follows:
\begin{equation}
  v_{g} = 0.0016(-T)^{1.4471}, \quad T<0^{\circ}\text{C}
\end{equation}

where \(T\) is the supercooling degree, and the unit is \(^{\circ}C\).

In the experiment, silica nanoparticles were introduced as nucleating agents, significantly reducing the nucleation energy barrier and promoting the nucleation process at low undercooling. The profile of the droplet's ice front primarily depends on the shape of the initial ice nucleus and the subsequent growth of the ice crystals. In the experiment, the ice front of the droplet did not undergo complex deformation, suggesting that the diversity of the droplet’s ice front may depend on the shape of the initial ice nucleus. For ice crystal growth in pure water, the growth velocity is mainly determined by the latent heat release at the solid-liquid interface (\cite{glicksman2004dendritic}). Therefore, under a constant ice front growth velocity, the stability of the latent heat release at the ice front interface in the experiment is indicated.

\subsection{Accumulated deformation of droplets in the freezing process}
\begin{figure*}[b!]
\centering
\includegraphics[width=0.9\linewidth]{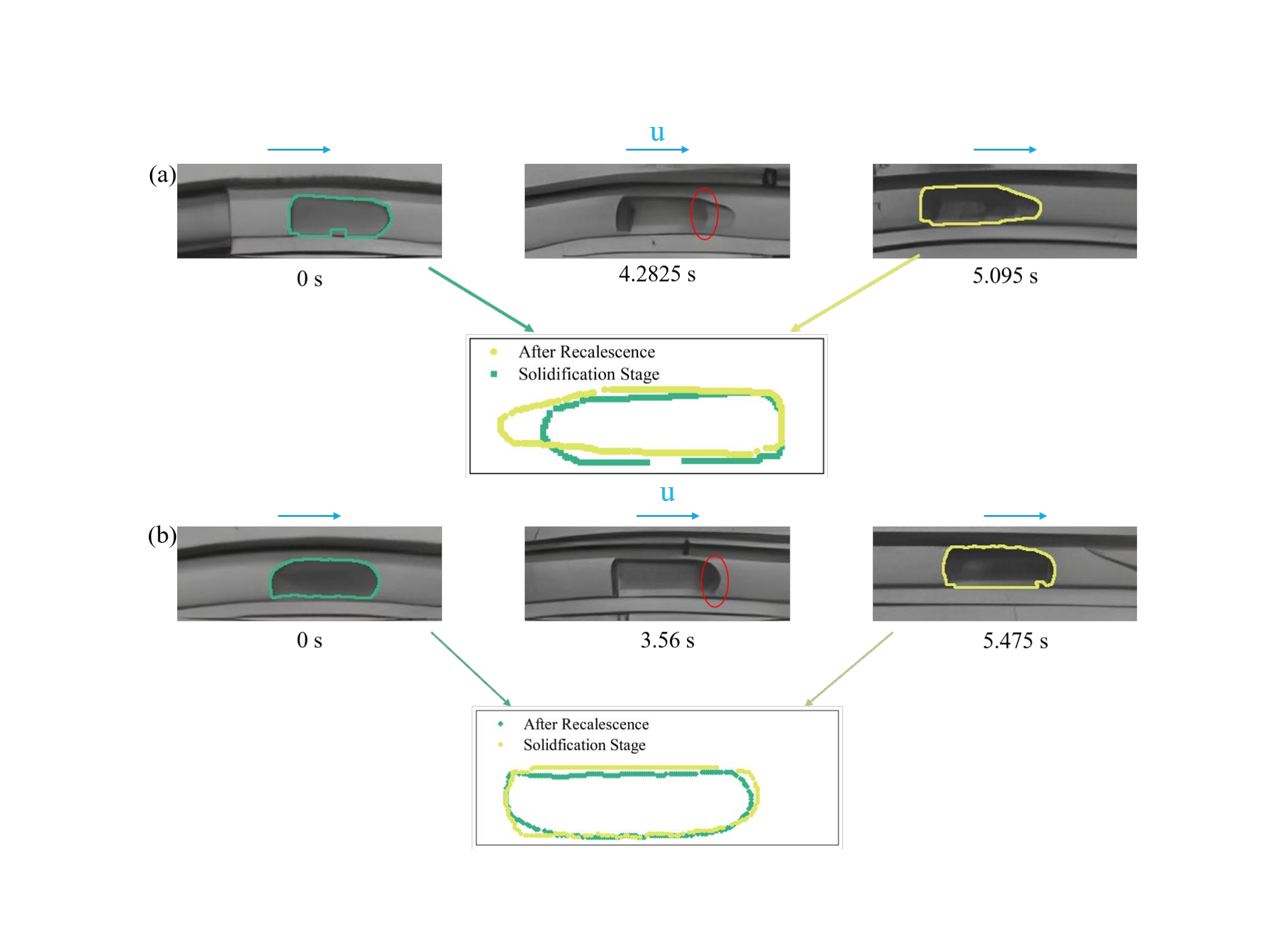}
\caption{Deformation of a droplet during solidification. (a) Deformed droplet; (b) Undeformed droplet.}
\label{fig:9}
\end{figure*}
During the freezing process, the droplet undergoes continuous deformation. This deformation is not very noticeable during the recalescence phase but becomes significantly pronounced during the solidification phase. As shown in Figure \ref{fig:9}, two types of solidification processes in the confined pipeline are presented. Due to the higher density of water compared to ice, the droplet will experience volumetric expansion in the pipeline, which has been considered in other studies(\cite{sgro2007thermoelectric,tarn2021homogeneous,tarn2020chip,stan2009microfluidic,isenrich2022microfluidic}). We allowed the droplet to solidify in a confined pipeline, where the wall surface is hydrophobic and oleophilic, forming an oil film on the wall. As a result, the ice does not block the pipeline. Figure \ref{fig:9} shows the droplet contours obtained using background subtraction and contour extraction algorithms. We compared the droplet contours immediately after the recalescence stage with those after solidification. In Figure \ref{fig:9}(a), solidification begins at the waist of the droplet. In Figure \ref{fig:9}(b), solidification starts from the droplet's head and tail. This difference is likely due to the droplet maintaining a temperature of 0°C after the recalescence stage, while the surrounding environment is significantly below 0°C. The initial solidification points are located in regions with lower temperatures than elsewhere. In regions with lower temperatures, the droplet's interfacial tension is higher, while in warmer regions, the interfacial tension is lower. The differences in interfacial tension and thermal convection (\cite{chen2014study,wildeman2017fast}) together lead to droplet deformation. Variations in interfacial tension result in different surface energies across regions with varying temperatures, driving the droplet to undergo shape deformation. Meanwhile, thermal convection further influences the droplet's internal temperature distribution and heat transfer, exacerbating the deformation.

\section{Randomness and Controllability of Droplet Dynamic Solidification}
\begin{figure*}[bp]
\centering
\includegraphics[width=\linewidth]{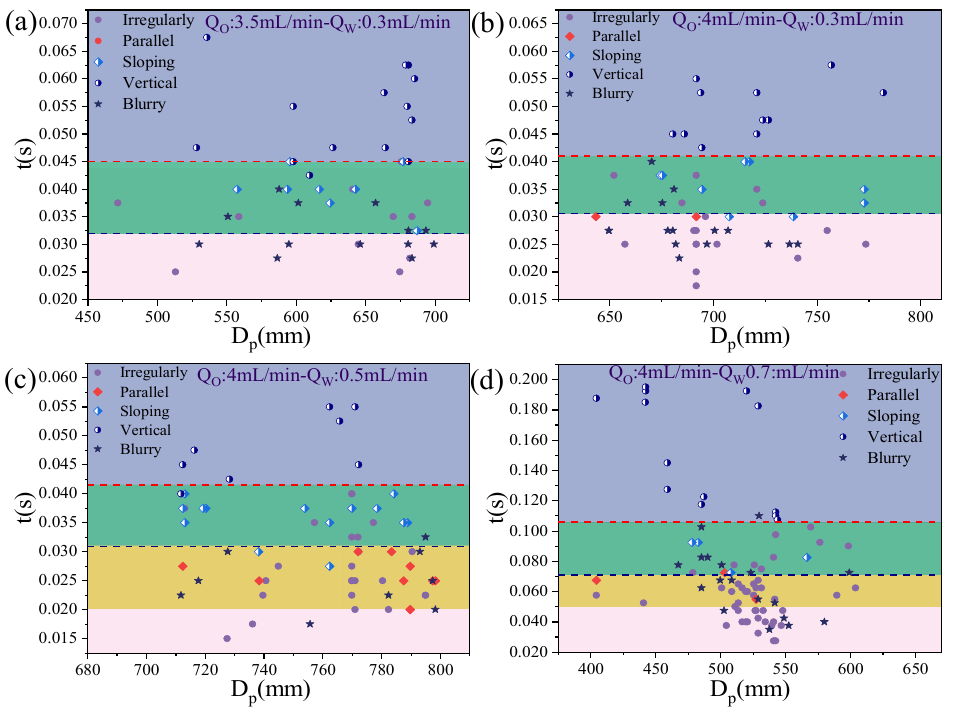}
\caption{Recalescence time and nucleation position distribution under different interface propagation patterns.}
\label{fig:10}
\end{figure*}

Figure \ref{fig:10} shows the distribution diagrams of the droplet recalescence time and the nucleation position under different flow rates and ice front profiles. 
It can be seen from the figure that the vertical ice front profile has the longest time in the recalescence stage, followed by the inclined one, and finally the parallel ice front profile. At the same time, the recalescence time of the irregular and blurred ones spans between the inclined and parallel ones. Under the same ice front propagation velocity, the farthest distance from the nucleation to the edge of the droplet determines the recalescence time. The recalescence time reflects the randomness of the nucleation position inside the droplet. For the same ice front profile, different nucleation positions will lead to different recalescence times. Therefore, the same ice front profile can have different recalescence times. Thus, Figure \ref{fig:10} shows the randomness of the nucleation position inside the droplet, and there is no relationship between the randomness and the ice front profile. In Figures \ref{fig:10}(a) and (b), droplets with ice front parallel to the pipeline axis are absent or rarely observed, whereas in Figures \ref{fig:10}(c) and (d), such droplets are abundant. This phenomenon may be related to the formation of ice nuclei, further highlighting the randomness of ice nucleation. The abscissa in Figure \ref{fig:10} represents the droplet nucleation position, that is, the position of the droplet in the pipeline when it is nucleated. It can be seen that regardless of the ice front profile, the nucleation positions of the droplets are not concentrated at a certain point but are scattered in the pipeline and occur randomly at different positions. 

\begin{figure*}[!htbp]
\centering
\includegraphics[width=\linewidth]{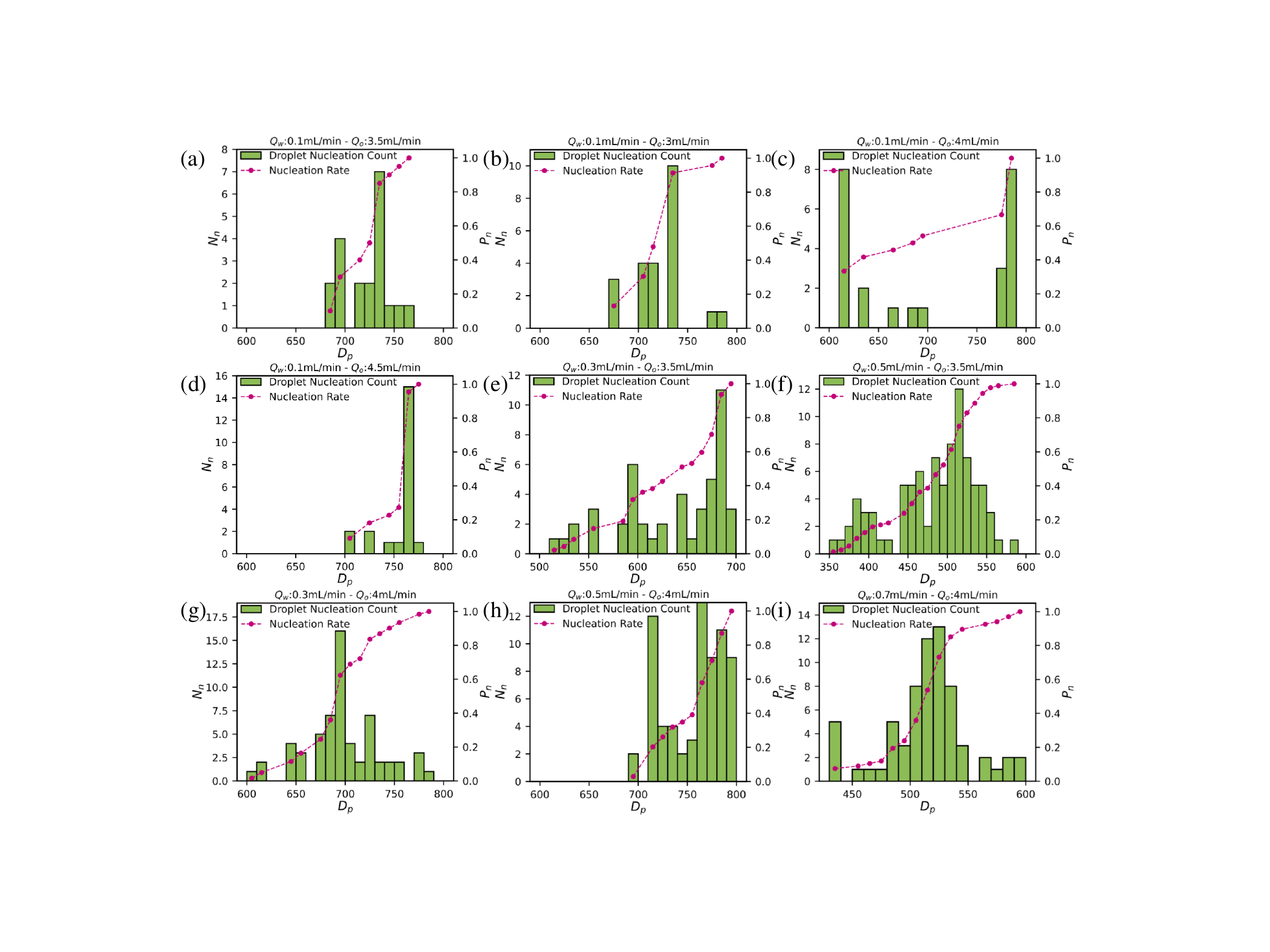}
\caption{Histogram of droplet nucleation positions within the channel and the cumulative ice nucleation rate (total number of nucleated droplets / total number of droplets) of droplets.}
\label{fig:11}
\end{figure*}

Figures \ref{fig:11}(a)-(i) are the statistical histograms of the nucleation positions of droplets in the pipeline under different flow rates and the cumulative nucleation rate of droplets (total number of nucleated droplets / total number of droplets). It can be seen from the figures that the droplet nucleation is relatively dispersed, that is, nucleation occurs at different positions, demonstrating the randomness of droplet nucleation within the pipeline. This randomness is the main factor affecting the digital solidification of droplets. However, in Figure 11, regardless of the flow rate, there will always be a position with the largest number of nucleations, and a large number of droplets will nucleate near this position. We analyzed the droplets by using statistical methods. Through statistical methods, we can understand the distribution law of droplet nucleation to a certain extent and provide references for further exploring the freezing behavior and control methods of droplets. 

The flatness factor is a statistic representing the kurtosis of data in statistics. It is used to describe the distribution pattern of data and reflects the sharpness of the peak of the data distribution, that is, the concentration degree of the data. The value of the flatness factor can help us understand the distribution characteristics of the data. When the flatness factor is relatively large, it indicates that the data has less dispersion and is more concentrated; when the flatness factor is relatively small, it means that the peak of the data distribution is relatively flat and the data is more dispersed. The defining formula of the flatness factor is as follows: 
\begin{equation}
    K=\frac{\mu^4}{\sigma^4}
\end{equation}

Among them, $\mu^4$ is the fourth-order central moment, and $\sigma$ is the standard deviation, which is defined by the following formulas: $\mu^{k} = \int_{-\infty}^{+\infty} (x - M_{x})^{k} p(x) \, dx$, $D_{x} = M(x - M_{x})^{2} = \int_{-\infty}^{+\infty} x^{k} p(x) \, dx$, $\sigma = \sqrt{D_{x}}$. $D_x$ is the variance, $\mu^k$ is the $k$-th order central moment, and in the flatness factor, $k = 4$. $M_x$ is the mathematical expectation, specifically: $M_{x} = \int_{-\infty}^{+\infty} x \cdot P(x) \, dx$. 

\begin{figure}[b!]
\centering
\includegraphics[width=\linewidth]{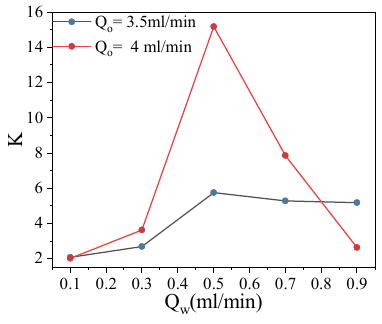}
\caption{The variation of the flatness factor with water-phase flow rate under different oil-phase flow rates.}
\label{fig:12}
\end{figure}

Figure \ref{fig:12} illustrates the variation of the flatness factor with water phase flow rate under different oil phase flow rates. When the oil phase flow rate is constant and the water phase flow rate is adjusted, the flatness factor initially increases and then decreases with increasing water phase flow rate, reaching its maximum value at a water phase flow rate of 0.5 ml/min. This indicates that in our experiments, a water-phase flow rate of 0.5 ml/min represents the optimal flow rate for droplet nucleation. When $Q_{o} = 3.5$ ml/min and $Q_{w} > 0.5$ ml/min, the flatness factor decreases slowly. This phenomenon can be attributed to the increased droplet frequency at $Q_{w} > 0.5$ ml/min, which shortens the distance between droplets. Since the droplets are generated as slug flow in the T-junction, post-recalescence droplets experience friction with the channel walls, leading to a reduction in droplet velocity. Droplets yet to nucleate then collide with nucleated droplets ahead of them, inducing nucleation due to perturbation. Such collision-induced nucleation, however, is not desired in our study.  Therefore, $Q_{w} = 0.5$ ml/min is identified as the optimal water phase flow rate, where droplet nucleation achieves maximum uniformity without inter-droplet disturbances. According to the 1 - $\sigma$ principle, 68.27\% of data in a normal distribution falls within the range [x - $\sigma$, x + $\sigma$]. Normal distributions have a kurtosis of 3, while distributions with kurtosis greater than 3 are classified as leptokurtic, indicating higher data concentration. Using the 1 - $\sigma$ principle, we quantified the droplet nucleation rate within the range [x - $\sigma$, x + $\sigma$].
The nucleation rate corresponds to the kurtosis, increasing initially with the water phase flow rate before decreasing. At $Q_{o} = 3.5$ ml/min and $Q_{w} = 0.5$ ml/min, the nucleation rate reached 79.2\%. When $Q_{o} = 4$ ml/min and $Q_{w} = 0.5$ ml/min, the nucleation rate is even as high as 97.1\%. Thus, the 1 - $\sigma$ method provides a straightforward and intuitive way to quantify the uniformity of droplet nucleation, offering technical support for the digitalization of droplet nucleation processes.

\section{Conclusion}

This study aims to explore the digitized freezing process of droplets in microchannels. To achieve this, a spiral-shaped milli-reactor was designed, and droplets were frozen using a thermostatic bath to control the temperature. The experiments revealed significant randomness in the droplet freezing process, primarily manifested in the random distribution of nucleation positions within the droplets. During the freezing process, the ice front exhibited five distinct patterns: blurry, irregular, parallel, sloped, and perpendicular. Observations indicated that the ice front maintained its initial shape during propagation and advanced at a constant velocity. However, the recalescence duration of droplets displayed noticeable variability, further corroborating the randomness of nucleation positions within the droplets.

In the solidification stage of the droplets, two solidification modes were identified: simultaneous solidification from both ends of the droplet and solidification initiating from the middle of the droplet. Solidification starting from the middle often led to droplet deformation, which may be a key factor affecting the effectiveness of digitized droplet freezing.

Moreover, the randomness of droplet nucleation in microchannels is a major factor influencing digitized freezing. To quantify and control this randomness, this study introduced the flatness factor, a measure of data dispersion. By varying the water phase (dispersed phase) flow rate, the flatness factor can be effectively adjusted, thereby impacting the digitization of droplet freezing. Under constant oil phase (continuous phase) flow rate conditions, the flatness factor initially increased and then decreased with changes in the water phase flow rate. The flatness factor reached its maximum value at a water phase flow rate of 0.5 ml/min, indicating the optimal digitization of droplet freezing at this flow rate. Thus, by adjusting the dispersed phase flow rate, the digitized freezing effect of droplets in microchannels can be effectively controlled.

In conclusion, this study provides experimental insights and analysis that serve as a reference for understanding the freezing process of droplets in microchannels. 
%------------------------------------------------

\begin{acknowledgements}
This article was funded by the National Key Research and Development Program (2017YFB0404503) and the Shanghai Natural Science Foundation (15ZR1416400).
\end{acknowledgements}

\section*{Appendix A}
This study employed the method proposed by Wheeler (\cite{wheeler1993computation}) to simulate the growth of ice dendrites in supercooled water. The phase-field equation and temperature equation governing the dendritic growth of a pure substance in a supercooled solution can be expressed as:
\begin{equation}
    \frac{\overline{\varepsilon}^{2}}{m}\frac{\partial\phi}{\partial t}=\phi(1-\phi)(\phi-\frac{1}{2}+30\overline{\varepsilon}\alpha\Omega u\phi(1-\phi))+\overline{\varepsilon}\nabla^{2}\phi 
\end{equation}
\begin{equation}
    \frac{\partial u}{\partial t}+\frac{1}{\Omega} p'(x) \frac{\partial \phi}{\partial t} = \nabla^2 u
\end{equation}

In the equations, $p(\phi)=\phi^{3}(10 - 15\phi + 6\phi^{2})$, $u$ is the dimensionless temperature, $\overline{\varepsilon}$ is the dimensionless interface thickness, and $m$ is the phase mobility. The relationships between $u$, $\overline{\varepsilon}$, $m$, $\alpha$ and physical parameters are as follows:
\begin{equation}
   u = \frac{T - T_m}{T_m - T_0}, \quad 
\alpha = \frac{\sqrt{2} w L^2}{12 c_p \sigma T_M}, \quad 
m = \frac{\mu \sigma T_m}{kL}, \quad 
\bar{\varepsilon} = \frac{\delta}{w} 
\end{equation}
\begin{figure*}[b!]
\centering
\includegraphics[width=\linewidth]{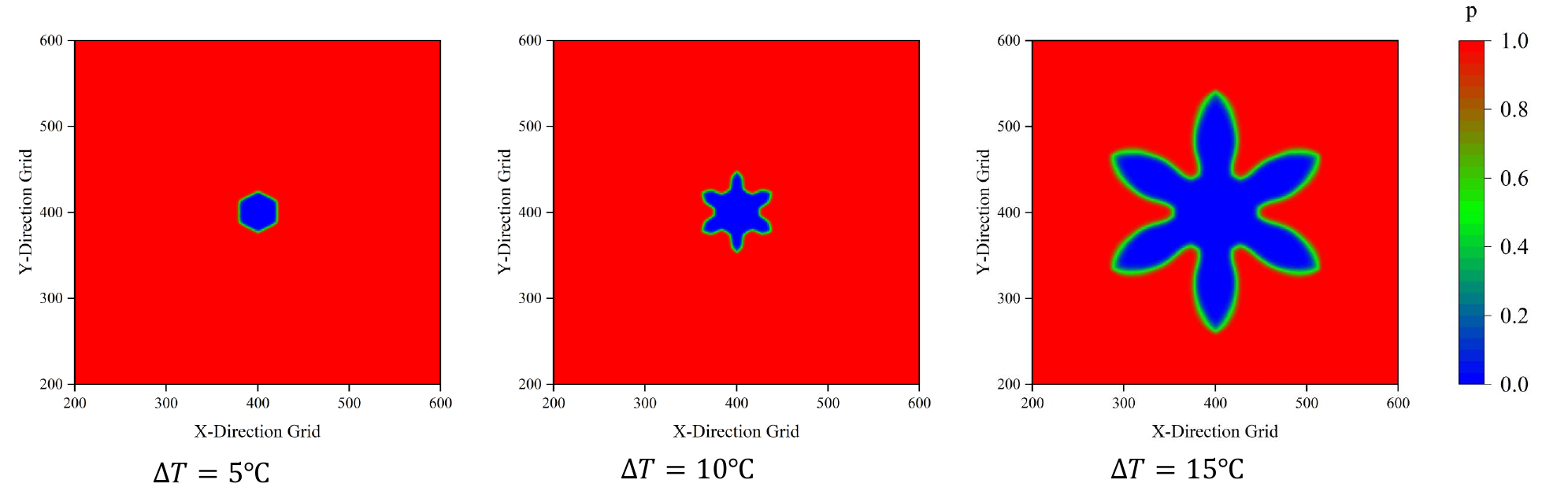}
\caption{The profiles of dendrites under different degrees of supercooling at the same time.}
\label{fig:13}
\end{figure*}

To describe the anisotropy of the interfacial free energy, the phase field parameter $\overline{\varepsilon}$ is considered to vary with the angle $\theta$ between the $<$100$>$ crystal phase and the interface direction within the (100) crystal plane, that is, $\overline{\varepsilon}=\overline{\varepsilon}f(\theta)$, where $f(\theta)$ is the anisotropy function. Therefore, for crystals with weak anisotropy, the parameter $\overline{\varepsilon}(\theta)$ related to the interfacial energy is introduced: 
\begin{equation}
    \overline{\varepsilon}\left(\theta\right)=\overline{\varepsilon}(1+\varepsilon_{k}\cos(j(\theta-\theta_0))
\end{equation}

In the equation, \(k\) is the modulus of anisotropy, and the value of \(k\) is taken as \(6\) in this paper; \(\varepsilon_{k}\) is the anisotropy strength. After considering the anisotropy, the Laplace operator in the phase field equation becomes: 
\begin{align}
\varepsilon^2 \nabla^2 \phi = \nabla^2_{(\theta)} \phi &= \nabla \cdot (\varepsilon^2(\theta) \nabla \phi) 
- \frac{\partial}{\partial x} \left( \varepsilon(\theta) \varepsilon'(\theta) \frac{\partial \phi}{\partial y} \right) \notag \\
&\quad + \frac{\partial}{\partial y} \left( \varepsilon(\theta) \varepsilon'(\theta) \frac{\partial \phi}{\partial x} \right)
\end{align}

In the actual solidification process, it is inevitable that the solid-liquid interface will be perturbed. The formation of secondary dendrites can be simulated by adding a perturbation term to the phase field governing equation. 
\begin{equation}
    A_n = -16 \phi^2 (1-\phi)^2 A_i R_n \times 30 \overline{\varepsilon} \alpha \Omega T \phi^2 (1-\phi)^2
\end{equation}

In the equation: $A_{i}$ represents the noise intensity, describing the magnitude of the introduced perturbation; $R_{n}$ is a random number in the range (-1,1), following a Gaussian distribution. The finite difference method is employed to solve the governing equations in this study. Time discretization uses first-order forward difference, spatial discretization uses central difference, and the Laplacian operator is discretized using a nine-point stencil scheme.

The initial conditions adopted in this paper are as follows: 
\begin{equation}
    x^2 + y^2 \leq r^2: \phi=0, T=0
\end{equation}
\begin{equation}
    x^2 + y^2 > r^2: \phi=1, T=-\Delta
\end{equation}

Both the phase field and the temperature field adopt zero-Neumann boundary conditions, that is $\partial \phi / \partial n = \partial T / \partial n = 0$.

The physical property parameters and computational model parameters used in this paper are (\cite{galfre2023phase}): \( T_m = 273.15\space K\), \(L = 305.65\space J/cm^3\), \(c_p = 4.212\space J/(cm^{3}K)\), \(\kappa = 0.0131\space cm^{2}/s\), \(\mu = 7.4\space cm/(Ks)\), \(\sigma = 2.85 \times 10^{-6}\space J/cm^2\), \(\omega = 2.1 \times 10^{-4}\space cm\), \(\alpha = 312.5\), \(m = 0.035\), \(\overline{\varepsilon} = 0.01\), \(\varepsilon_k = 0.06\), \(Q = 0.5\), \(R_0 = 10\). 

The number of grids used in this paper is \(800\times800\), with \(dx = dy = 0.02\) and \(dt = 5\times10^{-5}\). The above are dimensionless sizes, where \(dx = dx'/\omega\), \(dy = dy'/\omega\), and \(t = t'/(\omega^{2}/\kappa)\).

Figure \ref{fig:13} shows the profiles of dendrites at the same time under different degrees of supercooling. As the degree of supercooling increases, the growth velovity of dendrites also becomes faster. The growth velovity of dendrites mainly depends on the release of latent heat at the interface. Under a large degree of supercooling, the thermal gradient at the interface becomes thinner, so the growth velovity of dendrites becomes faster.

%----------------------------------------------------------------------------------------
%   REFERENCE LIST
%----------------------------------------------------------------------------------------
%\small
%\bibliographystyle{unsrt}%ieeetr}%apalike}
%\bibliographystyle{spbasic}
\bibliographystyle{microfluidics}
\bibliography{Experimental_studies_2024-12-1}

%\paragraph{Paragraph headings} Use paragraph headings as needed.
%\begin{equation}
%\end{equation}

%\begin{acknowledgements}
%\end{acknowledgements}

%\bibliography{}   % name your BibTeX database

% \begin{thebibliography}{}
% Author, Article title, Journal, Volume, page numbers (year)
% Author, Book title, page numbers. Publisher, place (year)
% etc
% \end{thebibliography}
\end{sloppypar}
\end{CJK}
\end{document}